\documentclass[twocolumn]{aastex63}

\shorttitle{Ring Morphology}
\shortauthors{Laune et al.}

\usepackage{xcolor}

\usepackage{amsmath}

\begin{document}

\title{Ring Morphology with Dust Coagulation in Protoplanetary Disks}

\author[0000-0002-0896-7393]{JT Laune}
\affiliation{Theoretical Division, Los Alamos National Laboratory,
  Los Alamos, NM 87545, USA}

\author[0000-0003-3556-6568]{Hui Li}
\affiliation{Theoretical Division, Los Alamos National Laboratory,
  Los Alamos, NM 87545, USA}

\author[0000-0002-4142-3080]{Shengtai Li}
\affiliation{Theoretical Division, Los Alamos National Laboratory,
  Los Alamos, NM 87545, USA}

\author[0000-0002-7329-9344]{Ya-Ping Li}
\affiliation{Theoretical Division, Los Alamos National Laboratory,
  Los Alamos, NM 87545, USA}

\author[0000-0001-6253-1630]{Levi G. Walls}
\affiliation{School of Physics and Astronomy, University of Minnesota,
  116 Church Street, Minneapolis, MN 55455, USA}

\author[0000-0002-1899-8783]{Tilman Birnstiel}
\affiliation{University Observatory, Faculty of Physics,
  Ludwig-Maximilians-Universit{\"a}t M{\"u}nchen, Scheinerstr.
  1, D-81679 Munich, Germany}

\author[0000-0002-9128-0305]{Joanna Dr\c{a}{\.z}kowska}
\affiliation{University Observatory, Faculty of Physics,
  Ludwig-Maximilians-Universit{\"a}t M{\"u}nchen, Scheinerstr.
  1, D-81679 Munich, Germany}

\author[0000-0002-1589-1796]{Sebastian Stammler}
\affiliation{University Observatory, Faculty of Physics,
  Ludwig-Maximilians-Universit{\"a}t M{\"u}nchen, Scheinerstr.
  1, D-81679 Munich, Germany}

\begin{abstract}

Tidal interactions between the embedded planets and their surrounding
protoplanetary disks are often postulated to produce the observed 
complex dust substructures, including rings, gaps, and asymmetries. 
In this Letter, we explore the consequences of dust coagulation 
on the dust dynamics and ring morphology. Coagulation of dust grains
leads to dust size growth which, under typical disk conditions, produces 
faster radial drifts, potentially threatening the dust ring formation. 
Utilizing 2D hydrodynamical 
simulations of protoplanetary
disks which include a full treatment of dust coagulation, we find
that if the planet
does not open a gap quickly enough, the formation of an inner ring is
impeded due to dust coagulation and subsequent radial drift. 
Furthermore, we find that
a ``buildup'' of sub-mm sized grains often appears in the
dust emission at the outer edge of the dust disk. 

\end{abstract}

\keywords{accretion, accretion disks ---
  protoplanetary disks ---
  planet–disk interactions ---
  planets and satellites: rings ---
  submillimeter: planetary systems}

\section{Introduction}
The high-resolution observations of protoplanetary disks (PPDs) by
ALMA have revealed some of the most stunning structures yet
\citep[e.g.,][]{andrews_disk_2018,Huang_2018ApJ...869L..42H,isella_disk_2018,macias_characterization_2019}.  Many disks exhibit
a system of rings and gaps in the dust emission.  Several mechanisms
have been proposed for creating rings of dust in protoplanetary disks;
one of the most commonly accepted scenarios is that embedded
protoplanets interact tidally with the disk, opening up a gap and
creating pressure bumps which effectively trap dust into
rings. Recently, \citet{Dong2017,Dong_2018ApJ...866..110D} found that
a single super-Earth sized planet may open up several rings and gaps
within its own orbit.  However, in order to trap dust within the
planet's orbit, the planet must compete with dust loss from radial
drift in the inner disk.

In a typical protoplanetary disk, 
the gas drag on dust grains 
tends to cause dust to drift towards the star
\citep{whipple_certain_1972,Weidenschilling1977,brauer_coagulation_2008,birnstiel_simple_2012}. 
This can be formulated in terms of the Stokes number of
the grain \citep{birnstiel_simple_2012},
$\mathrm{St} = \pi a \rho_p/(2\Sigma_g)$,
where \(a\) is the dust grain size, \(\rho_p\) its internal
density, and \(\Sigma_g\) the gas surface density. 
The radial drift timescale is given by
\begin{align}\label{eq:rdrift}
\tau_{\rm drift} \approx r/|v_r| = \frac{\mathrm{St}+\mathrm{St}^{-1}}{\Omega_K}
\frac{\Sigma_g v_K^2}{\gamma P}~~,
\end{align}
where $P$ is the vertically integrated gas pressure, $\approx~\Sigma_g c_s^2$,
and $\gamma =|\rm d\ln P/d\ln r|$.
For typical disks, using $\gamma = 1.41$, $c_s/v_K = 0.05$, 
and $\mathrm{St} = 1$ (which gives the largest drift velocity),
the drift timescale is about $90$ orbits. 

Due to the computational expense 
as well as physics uncertainties, dust coagulation is often neglected and 
single-species or multi-species approaches are used with pre-specified, fixed
distributions \citep{Fu2014a,Dong2017,Ricci2018,Miranda2016,zhang_disk_2018}. However, as dust grains grow in size, their Stokes numbers increase according to $\mathrm{St}$ and drift timescale shortens according to Equation~\ref{eq:rdrift}. 
Simulations which neglect coagulation
may therefore underestimate 
the effects of radial drift, especially if the initial
dust size is small.
Unless there is a feature blocking the radial drift, large dust
grains are typically lost as they travel inwards and accrete onto the
star. Hence, including dust coagulation may change the outcome of ring
formation dramatically.

Turbulent velocities are postulated to be the main source of
collisions for dust grains, and these velocities also grow with
the Stokes number. Once turbulent velocities exceed the fragmentation
velocities of the grains, large grain growth will halt as the
particles fragment into smaller pieces. For Stokes numbers \(\leq 1\),
we have the following maximum size to which dust particles grow, beyond 
which the dust size distribution falls exponentially
\citep{birnstiel_simple_2012,pinilla_trapping_2012,li_effects_2019}:
\begin{align}\label{eq:amax} 
a_{\rm max} 
= \frac{4\Sigma_g}{3\pi\alpha_{\rm vis}\rho_p} \frac{v_f^2}{c_s^2} ~~,
\end{align}
where $\alpha_{\rm vis}$ is the viscosity parameter for coagulation,
$v_f$ is the fragmentation velocity for the dust, and $c_s$ is the
sound speed in the gas. For our studies here, we use $v_f = 10$ m/s.
This implies that $a_{\rm max} \sim 8$ cm, if we use 
$\Sigma_g = 20 \mathrm{~g/cm^2}$, $\alpha_{\rm vis} = 10^{-3}$, 
$\rho_p = 1.2\mathrm{~g/cm^3}$, and $c_s = 300 \mathrm{~m/s}$.
This fragmentation limit on particle size is strict,
if the initial dust size is much smaller than 
$a_{rm max}$. 

There is much uncertainty about the coagulation timescale in
protoplanetary disks \citep{Brauer2007}, as the typical micron-sized
dust grains from the ISM gradually grow in size inside the PPD.  We
adopt the approach of \citet{birnstiel_simple_2012}, which derives the
particle growth timescale as approximately
$\tau_{\rm growth} \approx 1/( \epsilon_0 \Omega_K)$, where
\(\epsilon_0\) is the initial dust to gas ratio in the disk
\citep{brauer_coagulation_2008, ormel_closed-form_2007,
  youdin_particle_2007}.  The coagulation timescale to the maximum
fragmentation size \(a_{\rm max}\), then, is given by
\begin{equation}\label{eq:tcoag}
\tau_{\rm coag} = \tau_{\rm growth} \ln (a_{\rm max}/{a_0})~~,
\end{equation}
which is about $180$ orbits if we use $\epsilon_0 = 0.01$, $a_0 = 1$~\(\mu\)\text{m}
and \(a_{\rm max}\) from Equation \ref{eq:amax}.

By considering when \(\tau_{\rm drift} \approx \tau_{\rm coag}\),
\citet{birnstiel_simple_2012} derives the upper limit on dust grain
size due to losses from radial drift to be
\begin{align}\label{eq:adrift}
a_{\rm drift} = 0.55\times\frac{2\Sigma_d v_K^2}{\pi \gamma \rho_d c_s^2}~~,
\end{align}
where $\Sigma_d$ is the dust surface density, 
and the prefactor is tuned by simulations.
Putting in the typical values, we get $a_{\rm drift} \sim 5.5$ cm, which 
is slightly smaller than $a_{\rm max}$. 

When a planet is embedded in a PPD, it tries to open a gap in the disk. 
In order to estimate the gap-opening timescale, we adopt
the equation from \citet{lin_tidal_1986}, which neglects the effects of 
viscous stress:
\begin{align}\label{eq:tgap}
\tau_{\rm gap} \approx \frac{1}{\Omega_K}\left(\frac{M_*}{M_p}\right)^2
\left(2H_0\right)^5~~.
\end{align}
In this equation, \(H_0\) is the aspect ratio of the disk at the
planet's radius and \(M_p\) is the mass of the planet.  The viscous
timescale is typically much longer than the gap-opening timescale,
especially for the nearly inviscid disks in this work, and so for the
purposes of our study it is safe to ignore viscous stress. For a
planet mass of $M_p = 24\mathrm{~M_\oplus}$ around one solar mass
star, with $H_0 = 0.05$, Equation \ref{eq:tgap} implies a
gap opening timescale of around 300~orbits. Using the characteristic
parameters from Equations \ref{eq:rdrift} and \ref{eq:tcoag}, we see that
$\tau_{\rm coag} + \tau_{\rm drift}\approx 270$~orbits,
similar to $\tau_{\rm gap}$. 
Hence, we expect
a significant dust fraction will
coagulate and drift inwards from the planet's radius
before it opens a gap.
(Note that the combined effect of
coagulation and drift is more complicated than adding these two timescales.)

In the inner regions of a PPD, the dust coagulation
and subsequent loss due radial drift competes with the gap forming
process of an embedded planet.  
Due to coagulation, dust grains in the inner disk grow to their
fragmentation-limited size \(a_{\rm max}\) over a time \(\tau_{\rm coag}\).  
After reaching size  \(\sim a_{\rm max}\), 
the large grains drift inwards
over a time \(\tau_{\rm drift}\) if no pressure trap has formed. Hence, if
\(\tau_{\rm gap} > \tau_{\rm coag}+\tau_{\rm drift}\), we do not expect a ring of
dust to form inside the planet's orbit, and vice versa.  

In this Letter, we explore how the gap opening
process (which is determined by \(M_p/M_*\) and \(H_0\)) affects the
ring formation whenever full dust coagulation
is included in the disk's evolution. In \S 2, we will discuss our
simulation methods and set-ups. In \S 3, we present our main results,
followed by a summary and discussion in \S 4. 

\section{Methods}
We perform 2-D hydrodynamic simulations coupled with a
full coagulation treatment of the dust size evolution in a model PPD.
We then 
post-process synthetic dust continuum images to analyze the
observational signatures of dust coagulation.

\subsection{Hydrodynamic model}
Hydrodynamic simulations of the protoplanetary disks were performed
with \mbox{LA-COMPASS} code
\citep{Li_2005ApJ...624.1003L,Li_2009ApJ...690L..52L}.
We choose an isothermal sound speed profile of
\begin{equation}
    \frac{c_s}{v_k} = \frac{H}{r} = H_0\left(\frac{r}{r_p}\right)^{1/4},
\end{equation}
where $H$ is the scale height of the gas and 
$r_p$ is the planet's radius, which we set to 30~$\mathrm{au}$.
At this radius, 1000 orbits corresponds to approximately 150,000 years. 
This profile corresponds to a locally isothermal temperature profile 
which scales as $T\propto r^{-1/2}$.
We note that \citet{miranda_planetary_2019} recently showed that
an adiabatic equation of state with exponent $\gamma=1.001$
can have an effect on the gap opening process for planets.
For all of our simulations, the star has a mass of 1~\(M_{\odot}\),
and the viscosity parameter  \(\alpha = 5\times 10^{-5}\) using the
\mbox{\(\alpha\)--prescription}.
Our simulation
domain extends from 12~\(\mathrm{au}\) to 480~\(\mathrm{au}\),
with a grid of resolution of
\(n_r\times n_\phi = 1024\times 1024\).
We examined the numerical convergence
of our simulations using higher resolution
($\times2048$, $\times3072$)
single species runs. We found the inner ring
formation outcome to be the same 
for the single species runs below,
while the structure and peak density
of the outer ring could change with the resolution.
The
disk is initialized with an axisymmetric gas profile of
\begin{equation}
\Sigma_g(r) = \Sigma_0
\left(\frac{r}{r_c}\right)^{-\beta}\exp\left(-\left(\frac{r}{
r_c}\right)^{2-\beta}\right),
\end{equation} 
with \(\beta= 1.0\), \(\Sigma_0 = 12.3\) g/cm\(^2\), and \(r_c =
120~\text{au}\). The dust is set with an initial dust-to-gas
ratio of 1\%. These parameters lead to a total disk mass of around
38~\(M_J\). The planet masses vary between 10~\(M_\oplus\) and
50~\(M_\oplus\), and \(H_0\) varies between 0.03 and 0.07. 
We choose
\(M_p\) and \(H_0\) combinations which lie in the vicinity of the critical
region where \(\tau_{\rm gap}\approx \tau_{\rm coag}+\tau_{\rm drift}\). 
We grow the planet over a
period of 10 orbits at the beginning of the simulation.

\subsection{Coagulation model}
The gas and dust dynamics equations are as in \citet{Fu2014a}, and we
include dust feedback reaction on the gas.  The coagulation scheme is
described in \citet{li_effects_2019} and
\citet{drazkowska_including_2019}, using the model of
\citet{birnstiel_gas-_2010}.  In order to simulate dust dynamics and
coagulation, we use 151 species of dust logarithmically spaced between
1 \(\mu\)\text{m} up to 1m-sized boulders, which gives us 25 species
per size decade.  The dust distribution is initialized at a size of 1
\(\mu\text{m}\).   LA-COMPASS includes a fluid for each species of
dust, and to determine the evolution of the dust sizes we explicitly
integrate the Smoluchowski equation in each spatial cell
\citep{smoluchowski_drei_1916}.  Turbulence and radial drift are
considered as sources of relative velocities between the dust
particles.  The turbulence for dust is governed by a separate
viscosity parameter \(\alpha_{\rm vis}\), for which we choose a value
of \(10^{-3}\).
We note that
$\alpha_{\rm vis}\neq\alpha$ in our runs.  This is because
$\alpha_{\rm vis}$ represents the dust turbulent velocities at the
midplane, which may differ from the global viscous evolution of the
disk characterized by $\alpha$. These two 
can be different if disk accretion is not governed by the magnetorotational
instability \citep{carrera_planetesimal_2017}.
Collisions above 10 m/s result in fragmentation, and
those below result in coagulation. The fragmentation outcome results
in a distribution according to a power law in the mass \(m\) of the
fragments, \(n(m)\mathrm{d}m \propto m^{-1.83} \mathrm{d}m\)
\citep{brauer_coagulation_2008, birnstiel_gas-_2010}.

We handle the dust coagulation in each cell using an operator
splitting approach alongside the gas and dust hydrodynamics.  
To save computing time, we implement a sub-stepping
routine and calculate the coagulation outcomes every 50 timesteps.

\subsection{Radiative transfer}
To create synthetic dust emission images of our disk models, 
we utilize the radiative transfer code RADMC-3D of
\citet{dullemond_radmc-3d:_2012}. 
The detailed procedure is the same as described in
\citet{Huang_2018ApJ...867....3H} and \citet{li_effects_2019}.
For RADMC-3D, we reduce the grid
size in the \(r\) and \(\phi\) dimensions to \(n_r \times n_\theta \times
n_\phi = 300\times 40\times 100\).  The algorithm used to downsample
and extrapolate the dust density recovered over 94\% of the total dust
mass for all cases. In order to calculate the dust temperature for
emission, we utilized the thermal photon Monte Carlo capabilities of
RADMC-3D with 500 million photons. The grain-size dependent dust
opacity is adopted from \citet{Isella2009} and
\citet{Ricci2010}. The star is assumed to have a blackbody
temperature of 4200~\text{K}.

For all of the disk models, we set the observation distance to 100 pc
and simulated the dust emission at zero inclination at 1000 orbits.
The star is assumed to have a blackbody temperature of 4200~K
and a mass of 1~M$_{\odot}$.
We imaged the disk models at 890 (ALMA Band 7), 1330 (ALMA Band
6), and 3000 \(\mu\)\text{m} (ALMA Band 3). The emission map is convolved with a
Gaussian beam with FWHM 35 mas in order to simulate observation of the
disk, which is comparable to the resolution achieved with the recent DSHARP survey
\citep{andrews_disk_2018}.

\section{Results}
In order to illustrate the effects of planet mass and aspect ratio on
the formation of rings interior to the planet's orbit, we will study
three representative models in more detail with a combination of 
parameters \(M_p = 24, 40\) \(M_\oplus\) and \(H_0 = 0.04, 0.05\),
Additionally, for each of them
we ran a separate single species simulation in order to compare. 
We will refer to them as \texttt{M24H005C/S}, \texttt{M24H004C/S},
and \texttt{M40H005C/S}, respectively. 
(The letters \texttt{C} and \texttt{S} refer to coagulation and
single species, respectively.) 
We use the density-weighted 
average dust size in the inner ring from the coagulation runs 
as the characteristic sizes for single species runs.  The dust size is 
$1.34, 1.26, 2.23$~cm for 
  \texttt{M24H005S}, \texttt{M40H005S}, and 
  \texttt{M24H004S}, respectively.
Many other runs are also made to map out the parameter space. 
First, we will use the model
\texttt{M24H005C} to verify the analytical formulas for the
characteristic scales given in the introduction.

\subsection{Characteristic scales}
\label{sec:org44edfcb}
\begin{figure*}[h]
\centering
\includegraphics[width=0.75\linewidth]{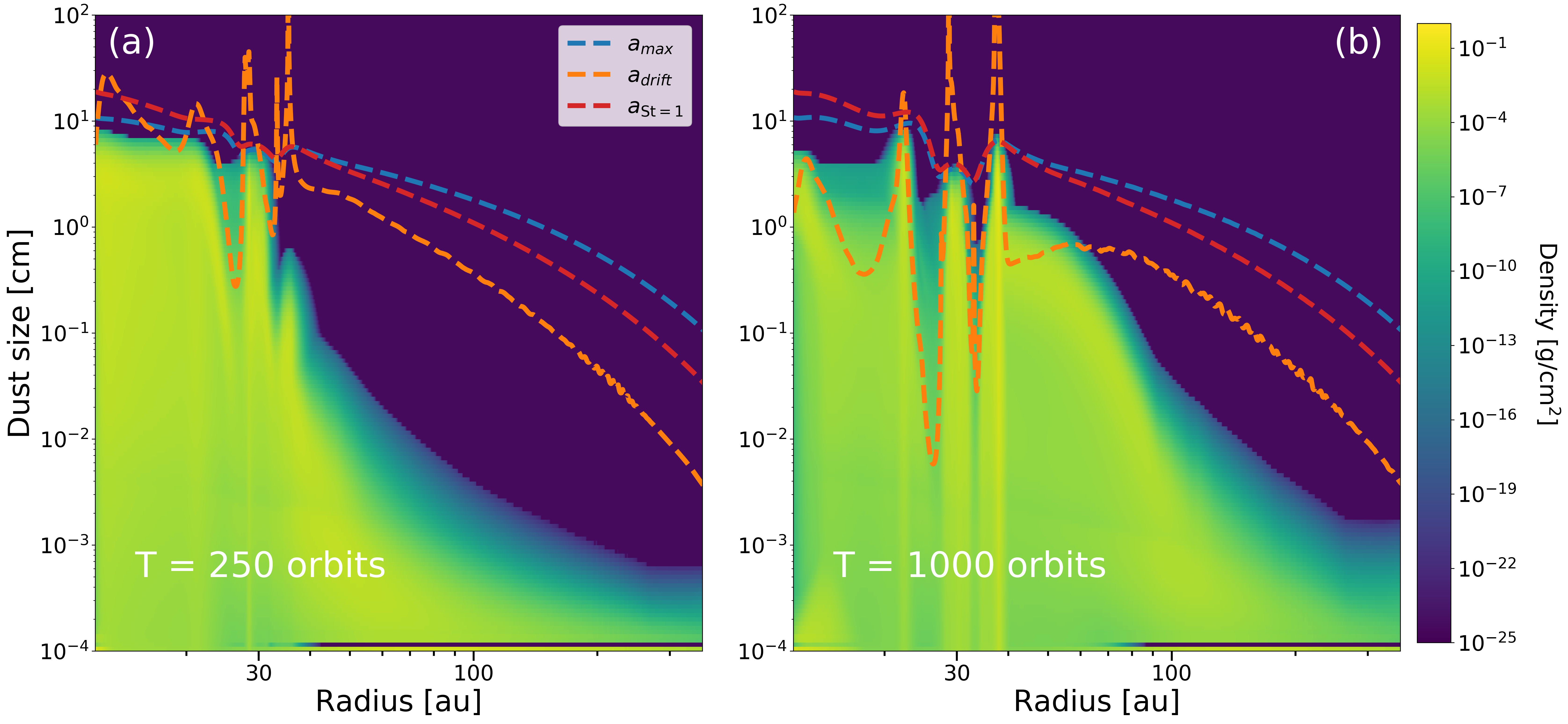}
\caption{
Results from \texttt{M24H005C}.
(a) Azimuthally averaged profile of the dust size
distribution at 250 orbits. The predicted $a_{\rm max}$ and $a_{\rm drift}$ 
are shown, along with the $a_{\rm St=1}$ curve. 
(b) Same as (a), but at 1000 orbits.
}
\label{fig:scales}
\end{figure*}

In Figure~\ref{fig:scales}, we have the radial dust size
distributions taken at 250 orbits and 1000 orbits (the last frame of
our simulation). Along with the size distributions, we have plotted
\(a_{\rm max}\) and \(a_{\rm drift}\). At 250 orbits, the dust in the
inner regions of the disk is approaching the fragmentation-limited
size \(a_{\rm max}\), while dust in the outer regions is still sub-mm.
At 1000 orbits, the particles have reached \(a_{\rm max}\) in the
rings, and they fit the shape of \(a_{\rm drift}\) well in the gaps.
The edge at $a_{\rm drift}$ is less sharp than $a_{\rm max}$ since the
radial drift growth limit is not as strict as the fragmentation limit.
This illustrates that in our simulations the coagulation, fragmentation 
and radial drift are all behaving close to expectation. 

\subsection{Full coagulation runs}
\begin{figure*}[h]
\centering
\includegraphics[width=0.75\linewidth]{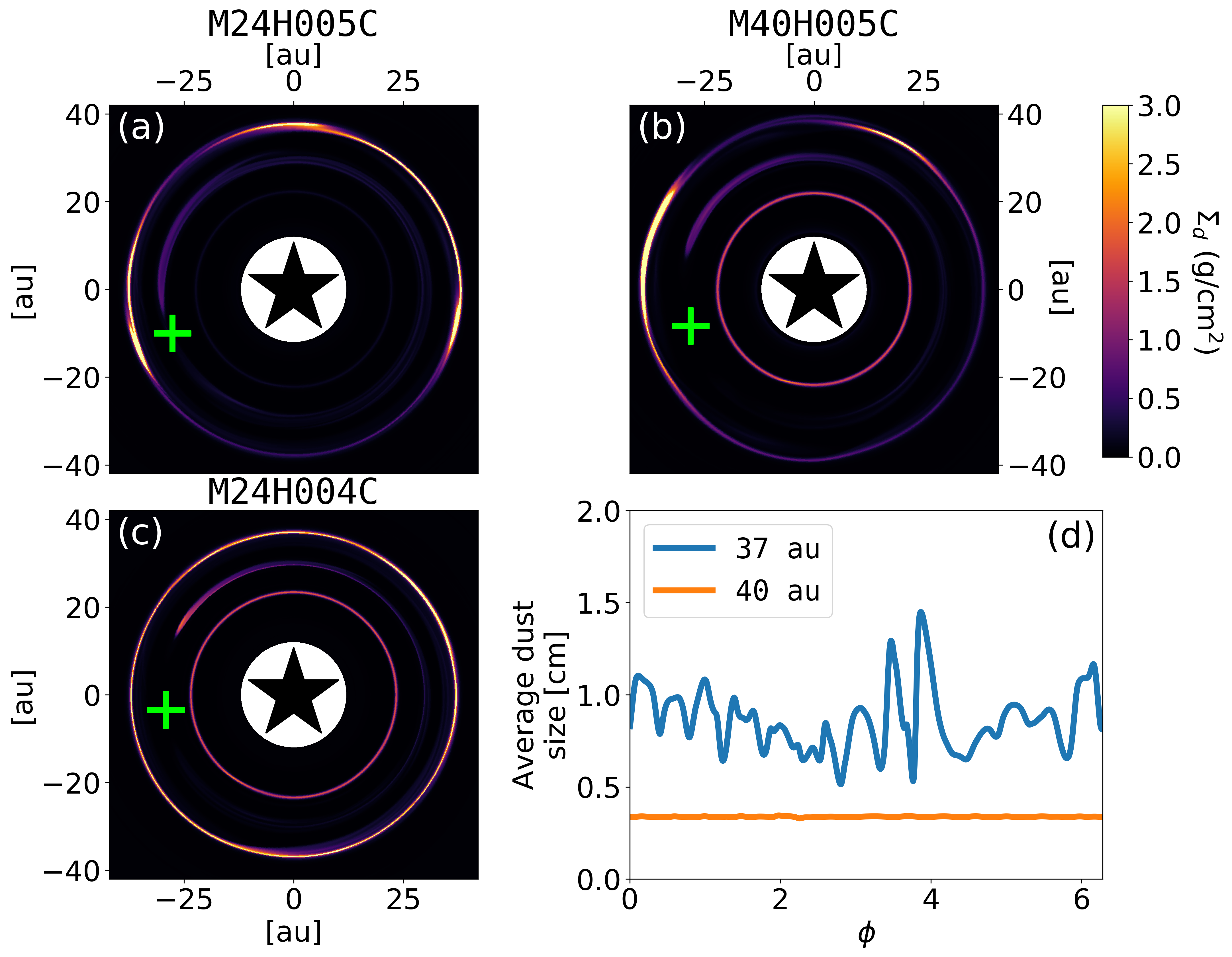}
\caption{
2-D dust density distributions and 1-D azimuthally averaged
dust radial profiles for the three representative runs. The green cross
denotes the location of the planet. All of these were taken at 1000 orbits.
(a) Dust surface density for \texttt{M24H005C}. The dust ring
inside the planet's orbits is greatly diminished compared to the others.
(b) Same as (a), but for \texttt{M40H005C}. Increasing the planet
mass has created a much stronger ring interior to the planet.
(c) Same as (b), but for for \texttt{M24H004C}. Decreasing
the disk aspect ratio similarly creates a much stronger inner ring.
(d) Density-averaged dust size for \texttt{M24H005C} in the outer ring ($r=37~\rm au$)
and outside the outer ring ($r=40~\rm au$). We see dust size enhancement inside the ring
where the dust density is higher along with variation of the average density 
in the $\phi$-direction.  
}
\label{fig:dens2d1d}
\end{figure*}

In Figure~\ref{fig:dens2d1d}, we have plotted the dust density for
each of the models with full coagulation. In
\texttt{M24H005C}, \(\tau_{\rm gap} > \tau_{\rm coag} + \tau_{\rm drift}\), and we
see that the innermost ring is very weak. However, increasing the
planet mass or decreasing \(H_0\) lowers \(\tau_{\rm gap}\), and we see that
both \(\texttt{M40H005C}\) and \(\texttt{M24H004C}\) form strong inner rings.
In \texttt{M24H005C}, an inner pressure maximum forms in
the gas, but the dust coagulates and drifts beforehand.

The density-averaged size of the dust within the outer ring at 37~au of \texttt{M24H005C}
is given in the bottom right of Figure~\ref{fig:dens2d1d}. For comparison,
we have also plotted the average size for a region just outside
the ring at 40~au. Inside the ring,
we see an enhancement of the dust size due to the fact that the dust density
is higher in this region. Size enhancement within forming rings boosts radial drift
and contributes to the diminished inner ring in \texttt{M24H005C}.
There are also variations in the average
size along the $\phi$-direction, which we expect to be time-dependent at this stage
of disk evolution. 

\subsection{Single species runs}
\begin{figure*}[h]
\centering
\includegraphics[width=1.0\linewidth]{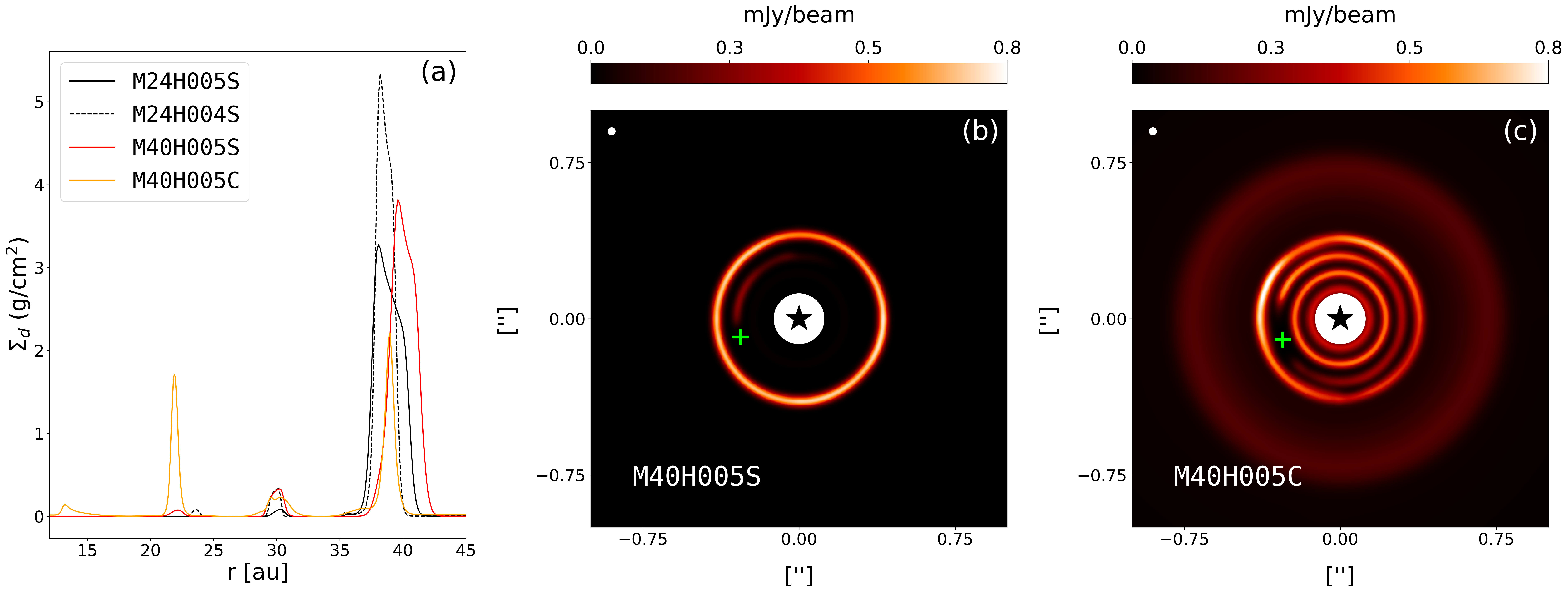}
\caption{
(a) Azimuthally averaged radial dust profiles for all three single
species runs at 1000 orbits. The profile from \texttt{M40H005C}
is also shown to demonstrate the effects from
coagulation.  (b) 1330~$\mu$m emission from the run
\texttt{M40H005S}  at 1000 orbits.  The
green cross and black star designate the planet's location and 
the star's location, respectively.  (c) Same as (b) but for
the run \texttt{M40H005C}.
}
\label{fig:dens1spec}
\end{figure*}

The dust profiles for the single species runs are presented in
Figure~\ref{fig:dens1spec}.
For all three runs, the ring within the planet's radius
either fails to form completely or is greatly diminished.
This is due to the fact that the dust is large from the
start of the simulation and can immediately drift quickly.
For these reasons, the single species dust emission plots in
Figure~\ref{fig:dens1spec} exhibit a much darker inner
disk than \texttt{M24H005C}.  One ring forms outside of the planet,
and only a portion of the ``horseshoe'' dust region at the planet's
radius is present. 
We also performed these single
species runs with a small dust size of 0.02~cm, and in each
case three rings formed. Because of the small size
and slow drift,
the rings were wider and the inner disk was brighter.

For comparison, 
\texttt{M24H005C} clearly exhibits four
distinct rings. Three immediately surround the planet's orbit, and a
fourth faint ring is located outside of these planetary gap rings.
The faint, outermost ring is from a buildup of sub-mm sized dust
particles which have grown from the \(\mu\)\text{m}-sized grains in
the outer disk but have not yet reached larger sizes where they drift
to the planet very quickly. As dust grains grow past 1~\(\mathrm{mm}\)
near this ring, they drift inwards and are quickly lost. This creates
the gap inside the ``buildup'' ring, since radial drift speed is
nonlinear. This feature is not present in \texttt{M40H005S}
due to the fact that the dust size is uniformly large near $\sim1$~cm.


\subsection{Varying aspect ratio and planet mass}
Using the same method of comparing peak dust density values between the
inner and outer rings as above, we have compiled the
ratios for all of our simulations after 1000 orbits into
Figure~\ref{fig:paramplot}.

We find good agreement between analytical predictions and our
simulation results. We used \(a_{\rm max}\) as the characteristic size
for the critical region where
\(\tau_{\rm gap}\approx\tau_{\rm coag}+\tau_{\rm drift}\). Inside and
close to this line, we see a gradient between the two
regimes, where the rings partially exist but are weaker than the
region in the bottom right of the parameter space. If the planet gap
opens in approximately the same amount of time as the dust coagulates
and drifts inwards, we still expect some dust to be trapped in the
inner ring.

\subsection{Synthetic dust emission}
\begin{figure*}[h]
\centering
\includegraphics[width=0.75\linewidth]{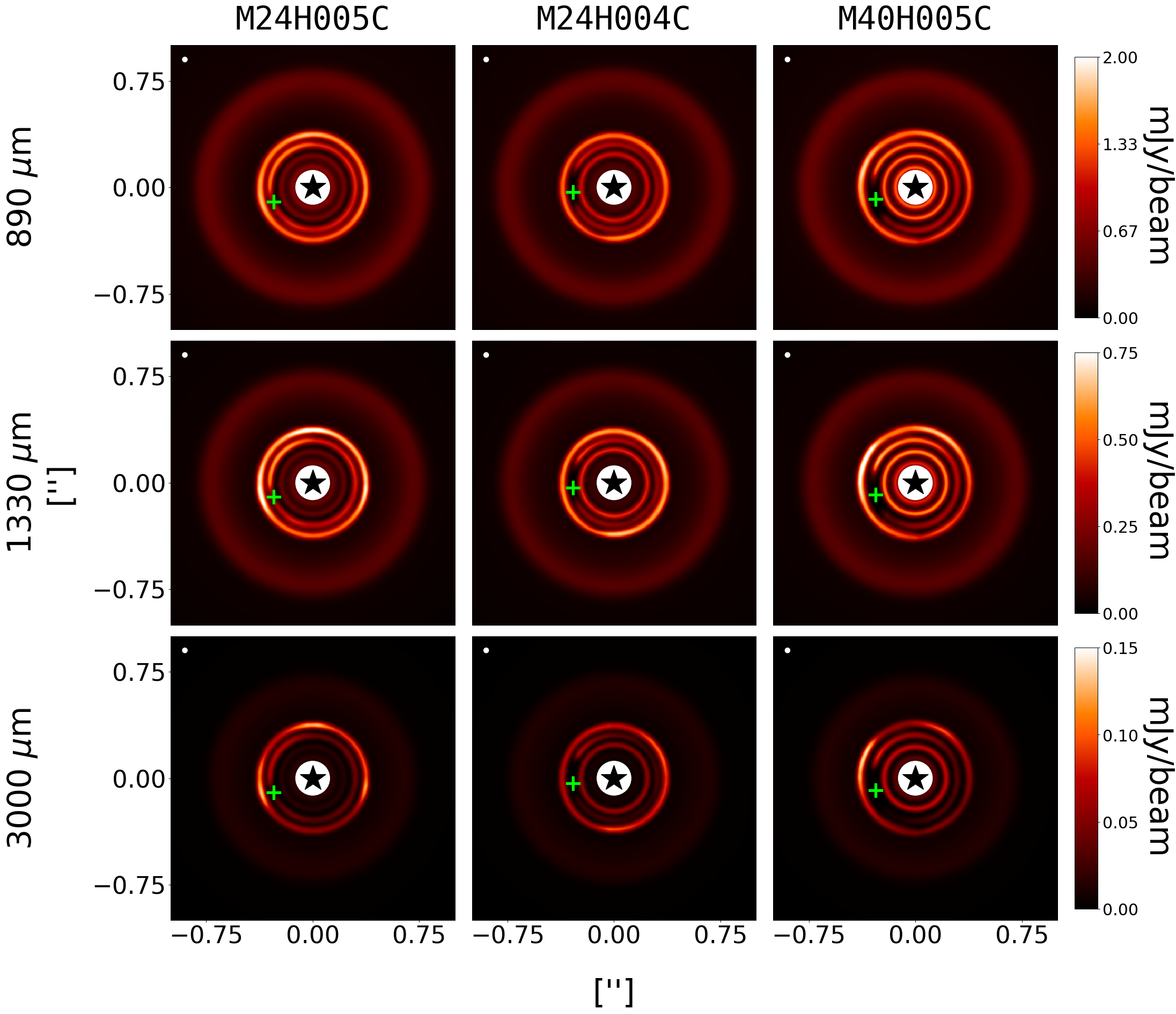}
\caption{
Simulated dust emission for wavelengths of 890 (ALMA Band 7), 1330
(ALMA Band 6), and 3000~$\mu$m (ALMA Band 3) for all three
representative full-coagulation models. The emission was 
calculated for a distance
of 100~pc. 
The green cross and black star indicate the planet and star locations,
as in Figure~\ref{fig:dens1spec}.
}
\label{fig:synemis}
\end{figure*}

The dust emission maps for our coagulation runs have been
plotted in Figure~\ref{fig:synemis}, for wavelengths of 890, 1330,
and 3000~\(\mu\)\text{m}. In all three wavelengths, the innermost ring
of \texttt{M24H005C} is much fainter than that of \texttt{M24H004C} or
\texttt{M40H005C}. The ``buildup'' outermost dust ring is present in all
three wavelengths, but it is relatively strongest in
890~\(\mu\)\text{m}.

The emission maps in Figure~\ref{fig:synemis} all exhibit
asymmetries in the outer ring adjacent to the planet, due to the
asymmetrical dust profile in this region. These asymmetries are most apparent
in the longer wavelengths, corresponding to larger dust particles which
are less coupled to the gas. In the overdense regions of the outer ring,
elevated dust densities could precipitate more efficient particle growth.

\section{Discussion and Summary}
In this Letter, we have conducted hydrodynamic and dust coagulation
simulations of PPDs spanning planet masses from 10 to
50~\(M_\oplus\) and disk aspect ratios between 0.03 and 0.07 in order to
test our analytical predictions for the formation of a dust ring
interior to the planet's orbit. The results of our simulations matched
our predictions well, and we found that dust coagulation and
subsequent radial drift impedes the inner dust ring from forming fully
for high aspect ratios and low planet masses, when 
\(\tau_{\rm gap}>\tau_{\rm coag}+\tau_{\rm drift}\). 

The dust distributions in simulations which included
coagulation differed from equivalent single species runs in several
major ways:
First, the inner dust ring was saved from radial drift
for runs with high planet
masses and low aspect ratios. 
Second, there was an additional ``buildup'' ring outside of the rings
immediately surrounding the planet gap.
Third, the ring exterior to the planet's orbit tends to be asymmetric.

One possible explanation of the asymmetrical outer dust ring
could be that the planet is triggering the Rossby wave instability 
\citep{lovelace_rossby_1999,li_rossby_2000, Li_2001ApJ...551..874L}. 
The asymmetries in the dust emission were strongest in the longest wavelength.
Vortices would be populated by larger particles because dust
grains with high Stokes numbers move more quickly towards the
pressure maxima and the elevated gas densities there allow particles to grow
to larger sizes. 
The varying dust density around the outer ring changes the size distribution
of the dust, which affects the spectral index and opacity of the region.
These effects are exacerbated if the dust grows to larger sizes and
concentrates into vortices more effectively.

\begin{figure*}[h]
\centering
\includegraphics[width=0.75\linewidth]{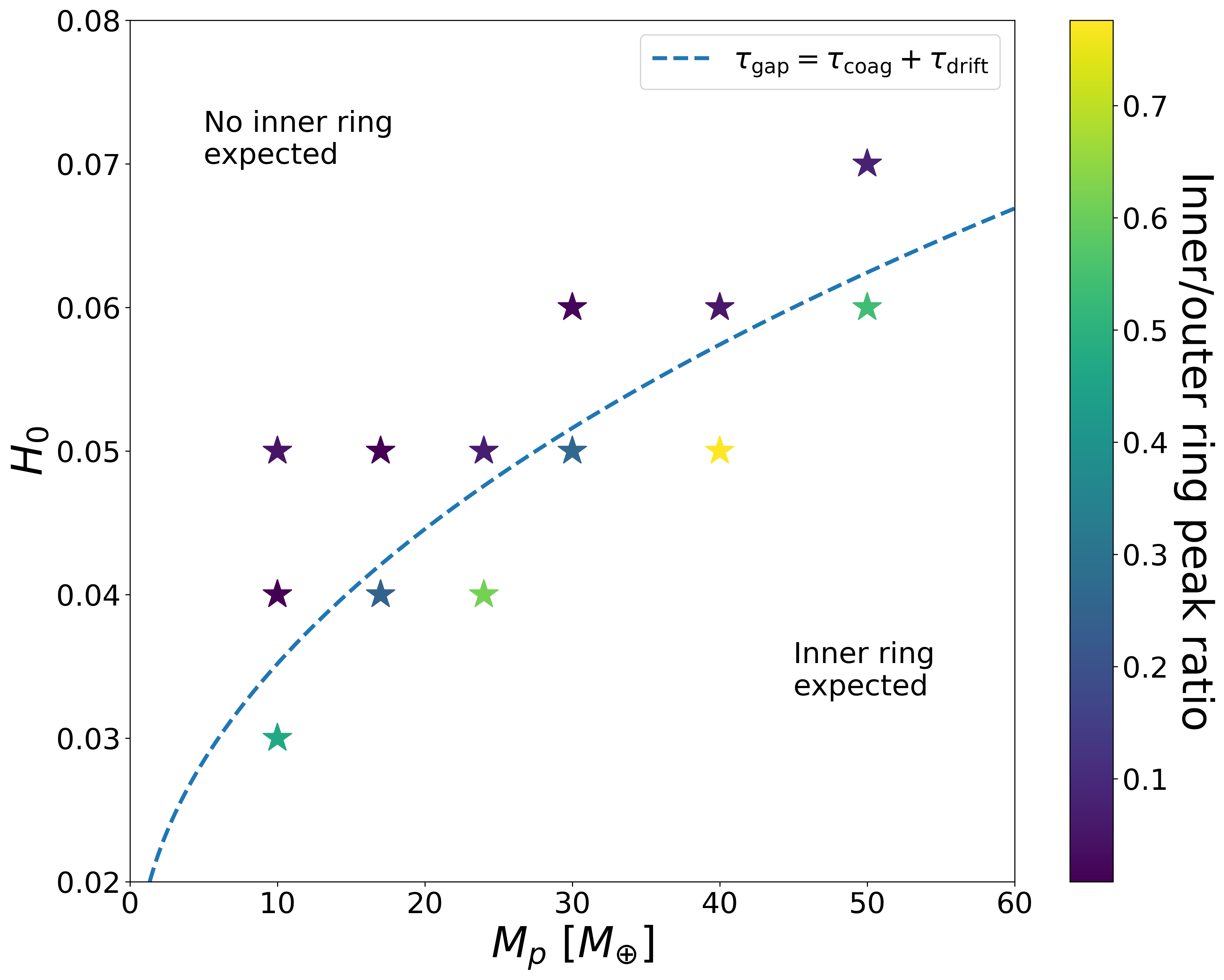}
\caption{The ratio between the outer ring's
peak density and the inner ring's peak density for all
full-coagulation runs performed.  The dashed line is
where $\tau_{\rm gap}\approx\tau_{\rm coag}+\tau_{\rm drift}$, with
$\tau_{\rm coag}$ and $\tau_{\rm drift}$ calculated using
$a_{\rm max}$.
}
\label{fig:paramplot}
\end{figure*}

There are several physical uncertainties in our coagulation model,
such as the fragmentation velocity and collision outcomes.
The size $a_{\rm max}$ is particularly sensitive
to the fragmentation velocity, with $a_{\rm max}\propto v_{\rm frag}^2$
from Equation~(\ref{eq:amax}). However, changes to the fragmentation
velocity would shift the boundary line for the parameter
space in Figure~(\ref{fig:paramplot}) but maintains its general behavior. 
Additionally, our assumption that the initial
dust size is uniform at 1~$\mu$m throughout the disk
is simplistic. In the future, more sophisticated initial dust size distributions
can be studied along with different planets masses and disk parameters. 
Given these uncertainties, 
our study is meant to highlight the interplay between dust growth, radial drift,
and ring formation in an approximate manner.

Our simulations neglected disk self-gravity (DSG), which can play an
important role in the gap-opening process of a disk.  In particular,
for some initial gas surface density values, including DSG may boost
the gap-opening process and decrease \(\tau_{\rm gap}\)
\citep{zhang_gap_2014}. This would effectively move the critical
band in Figure~\ref{fig:paramplot} upwards. In preliminary
LA-COMPASS simulations which include DSG, we do observe this effect,
as \(\tau_{\rm gap}\) decreases and a stronger inner ring forms for a run
equivalent to \texttt{M24H005C}. 


This Letter provides simulation evidence that
dust coagulation can impact the dust ring formation, brightness and 
asymmetries. They also
provide valuable inferences about the planet mass, disk aspect ratio, and
coagulation dynamics in a PPD system. 
These results motivate additional detailed studies on the physics
of dust coagulation in a more realistic PPD environment. 

We would like to thank the anonymous referee for helpful comments which improved the paper.
J.L., H.L., S.L., Y.P.L., and L.G.W. would like to gratefully acknowledge support
from LANL/CSES and NASA/ATP. 
T.B., J.D., and S.S. acknowledge funding from the European Research Council (ERC)
under the European Union's Horizon 2020 research and innovation
program under grant agreement No. 714769 and support from the
Deutsche Forschungsgemeinschaft (DFG, German Research Foundation)
Research Unit ``Transition Disks'' (FOR 2634/1, ER 685/8-1).
The numerical simulations in this study were performed
on LANL's Institutional Computing facilities.

\bibliography{main}{}
\bibliographystyle{aasjournal}

\end{document}